\documentclass{aa}
\usepackage{graphics}

\providecommand{\arcmin}{\ensuremath{^\prime}}
\providecommand{\arcsec}{\ensuremath{^{\prime\prime}}}

\newcommand{\mjysr}{\ensuremath{\mathrm{MJy\,sr^{-1}}}}
\newcommand{\fobs}{\ensuremath{F_\mathrm{Obs}}}
\newcommand{\fmodel}{\ensuremath{F_\mathrm{Model}}}

\begin{document}
\title{Solar System Objects in the ISOPHOT 170\,$\mu$m
       Serendipity Survey
       \thanks{Based on observations with ISO, an ESA project with
               instruments funded by ESA Member States (especially
               the PI countries: France, Germany, the Netherlands
               and the United Kingdom) and with the participation
               of ISAS and NASA.}}

\author{T.\ G.\ M\"{u}ller\inst{1,2} \and
        S.\ Hotzel\inst{3} \and
	M.\ Stickel\inst{3}}
	  
\authorrunning{M\"uller et al.}
\titlerunning{SSOs in the ISOSS}

\offprints{tmueller@mpe.mpg.de}

\institute{Max-Planck-Institut f\"ur extraterrestrische Physik,
	   Giessenbachstra{\ss}e, 85748 Garching, Germany
	   \and
	   ISO Data Centre, Astrophysics Division, Space Science
	   Department of ESA, Villafranca, P.O.\ Box 50727,
	   28080 Madrid, Spain (until Dec.\ 2001)
	   \and
	   ISOPHOT Data Centre, Max-Planck-Institut f\"ur Astronomie,
	   K\"onigstuhl 17, 69117 Heidelberg, Germany}

\date{Received / Accepted; compilation date: \today }

\abstract{
  The ISOPHOT Serendipity Survey (ISOSS) covered approximately 15\,\%
  of the sky at a wavelength of 170\,$\mu$m while the ISO satellite
  was slewing from one target to the next. By chance ISOSS slews
  went over many solar system objects (SSOs). We identified
  the comets, asteroids and planets in the slews through a fast
  and effective search procedure based on N-body ephemeris and
  flux estimates. The detections were analysed from a calibration and
  scientific point of view.
  Through the measurements of the well-known asteroids Ceres, Pallas,
  Juno and Vesta and the planets Uranus and Neptune it was possible to
  improve the photometric calibration of ISOSS and to extend it to
  higher flux regimes. We were also able to establish calibration
  schemes for the important slew end data.
  For the other asteroids we derived radiometric diameters and albedos
  through a recent thermophysical model. The scientific results 
  are discussed in the context of our current knowledge of
  size, shape and albedos, derived from IRAS observations, 
  occultation measurements and lightcurve inversion techniques.
  In all cases where IRAS observations were available we confirm
  the derived diameters and albedos. For the five asteroids without
  IRAS detections only one was clearly detected and the radiometric
  results agreed with sizes given by occultation and HST observations.
  Four different comets have clearly been detected at 170\,$\mu$m
  and two have marginal detections. The observational results are
  presented to be used by thermal comet models in the future.
  The nine ISOSS slews over \object{Hale-Bopp} revealed extended
  and asymmetric structures related to the dust tail. We attribute
  the enhanced emission in post-perihelion observations to large
  particles around the nucleus. The signal patterns are indicative of
  a concentration of the particles in trail direction.
}
\maketitle

\section{Introduction} \label{sec:int}

  The Infrared Space Observatory (ISO) (Kessler et al. \cite{kessler96}) made
  during its lifetime between 1995 and 1998 more than 30\,000 individual
  observations, ranging from objects in our own solar system
  right out to the most distant extragalactic sources. The solar
  system programme consisted of many spectroscopic and photometric
  studies of comets, asteroids, planets and their satellites at 
  near- and mid-infrared (near-/mid-IR) wavelengths
  between 2.5 and 45\,$\mu$m. At far-infrared (far-IR) wavelengths,
  beyond 45\,$\mu$m, the programmes were limited to spectroscopic
  observations of the outer planets, 3 satellites (Ganymede, Callisto,
  Titan), 3 comets (P/Hale-Bopp, P/Kopff, P/Hartley 2) and 4 asteroids
  (Ceres, Pallas, Vesta, Hygiea). Far-IR photometry on solar system
  objects was mainly done for calibration purposes (Uranus, Neptune and
  a few asteroids) and scientific studies of extended sources
  (P/Hale-Bopp, P/Kopff, P/Wild 2, P/Schwassmann-Wachmann, Chiron, Pholus).

  Additionally to the dedicated programmes on individual sources, ISO
  also made parallel and serendipitous observations of the sky. ISOCAM
  observed the sky in parallel mode a few arc minutes away from the
  primary target at wavelengths between 6 and 15\,$\mu$m (Siebenmorgen
  et al. \cite{siebenmorgen00}). The ISOPHOT Serendipity Survey
  (Bogun et al. \cite{bogun96})
  recorded the 170\,$\mu$m sky brightness when the satellite was
  slewing from one target to the next. LWS performed parallel
  (Lim et al. \cite{lim00}) and serendipitous surveys
  (Vivar\`{e}s et al. \cite{vivares00}) in the far-IR.
  These complementary surveys contain many interesting objects, but the
  scientific analysis only began recently. The two LWS surveys and the
  ISOCAM parallel survey just underwent a first data processing and
  could therefore not be considered in the following.

  Source extraction methods for the ISOPHOT Serendipity Survey (ISOSS)
  have been developed by Stickel et al.\
  (\cite{stickel98a}, \cite{stickel98b}) for point-sources
  and by Hotzel et al.\ (\cite{hotzel00}) for extended sources.
  First scientific results were published recently
  (Stickel et al. \cite{stickel00}; T\'{o}th et al. \cite{toth00};
  Hotzel et al. \cite{hotzel01}) and work is ongoing to produce further
  catalogues of
  Serendipity Survey sources. To facilitate the production of
  source lists it is necessary to identify and exclude all SSOs from
  the survey data. This catalogue cleaning aspect was
  one motivation for the following analysis, but there are also the
  calibration and scientific aspects of the SSO investigations:
  A few well known asteroids and planets like
  Uranus and Neptune provide the possibility to test and extend the
  photometric calibration of ISOSS to higher
  brightness levels (M\"uller \& Lagerros \cite{mueller98}).
  For asteroids with known diameters, the surface
  regolith properties can be derived from the emissivity behaviour
  in the far-IR where the wavelength is comparable to the
  grain size dimensions.
  Additionally, reliable far-IR fluxes of asteroids allow
  diameter and albedo determinations for the less well-known targets.
  For comets the far-IR information is useful for coma and tail
  modeling (Gr\"un et al. \cite{gruen01}). The close connection
  between large particles and far-IR thermal emission also allows
  further studies of trail formation and the important processes
  of dust supply for the interplanetary medium.

  In the following sections we present and discuss the ISOSS
  data with emphasis on solar system targets
  (Sect.~\ref{sec:obs}). This also includes the data processing,
  point-source extraction and calibration aspects. An iterative and fast
  method to search for SSOs in large sets of slewing data is explained
  in Sect.~\ref{sec:ext}.
  The encountered SSOs are then separated into 2 categories:\\
  1) Well known asteroids and the planets \object{Uranus} and
     \object{Neptune}, which were used to test and extend
     the photometric calibration of ISOSS
     (Sect.~\ref{sec:calres}).\\
  2) Asteroids and comets, for which the far-IR fluxes were used
     a) to derive diameters and albedos (asteroids) or
     b) to give a qualitative and quantitative description
        of the observational results for future modeling (comets)
     (Sect.~\ref{sec:scires}).\\
  In Sect.~\ref{sec:con} we summarize
  the results and give a short outlook to future projects.

\section{Observations and Data Reduction}
\label{sec:obs}

\begin{figure}[btp!]
  \resizebox{\hsize}{!}{\includegraphics{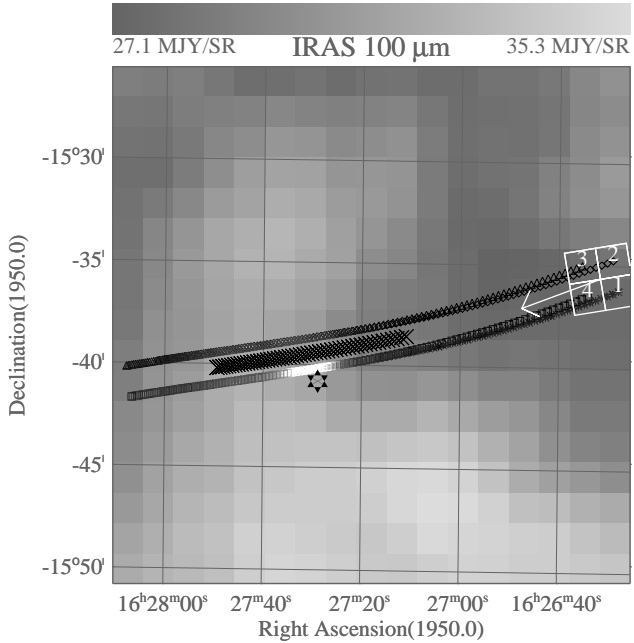}}
  \caption{IRAS 100\,$\mu$m map with the slew paths of the 4 C200 pixels 
	overplotted. The symbols used for the individual
	detector pixels are the same as in Fig.~\ref{fig:sso-in-slew2}.
	Here, their colour coding indicates qualitatively
	the measured intensities. The intensity peaks in Pixel~1 
	and 4 have no correspondence in the IRAS map.
	The star symbol marks the N-body ISO-centric position of \object{Ceres}
	at the time of the slew. Measurements within 5\arcmin\
	from the detector centre are marked here and in
	Fig.~\ref{fig:sso-in-slew2} as black crosses. 
	The detector and pixel apertures as well as the scan direction are
	indicated. The slew TDT number is 09380600 (see also
  Table~\ref{tbl:method2}).
	\label{fig:sso-in-slew1}}
  \end{figure}

\begin{figure}[btp!]
  \resizebox{\hsize}{!}{\includegraphics{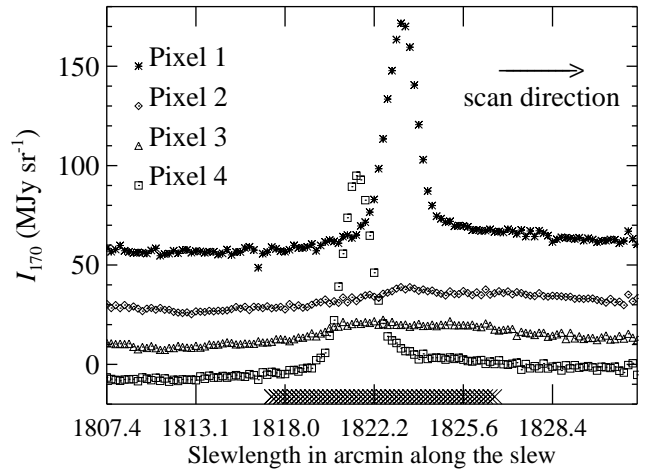}}
  \caption{The calibrated pixel intensities as a function
	of slew length. Intensities of Pixel~2--4 are shifted
	downwards in steps of 20~\mjysr. 
	The slew section shown corresponds to
	Fig.~\ref{fig:sso-in-slew1}. Black crosses mark measurements
	with detector positions closer than 5\arcmin\ to the calculated
	asteroid position.
	Pixel~1\&4 cross Ceres almost centrally.
	\label{fig:sso-in-slew2}}
  \end{figure}

  ISOPHOT Serendipity Survey (ISOSS) measurements were obtained with the
  C200 detector (Lemke et al. \cite{lemke96}), a 2$\times$2 pixel array of 
  stressed Ge:Ga with a pixel size of 89.4$^{\prime \prime}$.
  A broadband filter (C\_160) with a reference wavelength of
  170\,$\mu$m and a width of 89\,$\mu$m was used. The highest
  slewing speed of the satellite was 8$^{\prime}$/sec.
  During each 1/8\,sec integration time 4 detector
  readouts were taken, i.e.\ the maximum read out distance on the sky
  was 15$^{\prime \prime}$ yielding one brightness value per arcminute
  (see Figs. \ref{fig:sso-in-slew1} and \ref{fig:sso-in-slew2}). 
  During the ISO lifetime, about 550 hours of ISOSS
  measurements have been gathered, resulting in a sky
  coverage of approximately 15\%.


\subsection{Data Analysis\label{sub:data_analysis}}

  A standard data processing was applied using the ISOPHOT Interactive Analysis
  PIA\footnote{\footnotesize The ISOPHOT data presented in this paper were
  reduced using PIA, which is a joint development by the ESA Astrophysics
  Division and the ISOPHOT Consortium with the collaboration of the
  Infrared Processing and Analysis Center (IPAC). Contributing
  ISOPHOT Consortium institutes are DIAS, RAL, AIP, MPIK, and MPIA.}
  (Gabriel et al. \cite{gabriel97}) Version 7.2 software package.
  The detailed processing steps are given in Stickel et al.\
  (\cite{stickel00}).
  Special care had to be taken to correct gyro drifts between sequent
  guide star acquisitions of the star tracker.
  For point-sources, the deglitched and background subtracted signals of
  the 4 pixels were phase-shifted according to the position angle
  of the detector and co-added. Source candidates were searched for in this
  co-added stream by setting a cut of 3\,$\sigma$ of the local noise.
  Then, the source position perpendicular to the slew
  was determined from a comparison between signal ratios with a
  gaussian source model. The flux was afterwards derived from 2-D
  gaussian fitting with fixed offset position.

  In case of long slews, the surface brightness were derived from
  a measurement of the on-board Fine Calibration Source (FCS) preceding
  the slew observation. For short slews the default
  C200 calibration was used.
  To tie point-source fluxes derived from ISOSS
  to an absolute photometric level, dedicated photometric
  calibration measurements of 12 sources, repeatedly crossed with varying
  impact parameters were compared with raster maps on the same sources
  (Stickel et al. \cite{stickel98a}).
  The comparison between slew
  and mapping fluxes showed that for brighter sources the slewing
  observations miss some signal, probably due to transient effects in the
  detector output (Acosta-Pulido et al. \cite{acosta00}) in combination with
  detector non-linearities. For sources brighter than 30\,Jy Stickel et
  al.\ (\cite{stickel00}) found signal
  losses of 50\%, although the true losses were not well established due
  to a lack of reliable sources. For fainter sources ($<$10\,Jy) the flux
  loss in the slews is only 10--20\%.
 
\subsection{Source Extraction Methods}

  The SSOs were encountered at different slew speeds, 
  which can be characterized
  by `fast': above 3$^{\prime}$/sec;
  'moderate': 1.5$^{\prime}$/sec $<$ speed $<$ 3$^{\prime}$/sec;
  'slow': below 1.5$^{\prime}$/sec;
  'stop': at slewends, like a staring observation.
  But aspects like the background level, the detector history
  and impact parameters also play a crucial role in source
  extraction and flux calibration methods.

\subsubsection{Method 1}
\label{sec:sem_method1}

  Automatic point source extractor (Stickel et al. \cite{stickel00})
  for all slewing speeds above 1.5$^{\prime}$/sec and non-saturated
  crossings. All source candidates were cross correlated with the
  list of SSO candidates (see Fig.~\ref{fig:ssoextraction}) and
  the associations found carefully examined. Flux loss corrections
  (see Sect.~\ref{sub:data_analysis}) have to be applied.
	    
\subsubsection{Method 2}
\label{sec:sem_method2}
 
  For all slewing speeds, but using only the pixel with the highest
  signal and converting it to flux density as if the source was centred.
  This method leads to upper and lower flux limits only. The lower
  limits, designated by '$>$' or '$\gg$', are connected to clear detections.
  The quality of the lower estimate depends on the impact parameter.
  Useful upper limits have only been given for direct hits where no
  detection signal was seen. The upper limit then corresponds to the
  3$\sigma$-value of the background noise.
  Flux loss corrections (Sect.~\ref{sub:data_analysis}) have
  to be applied as for Method~1. 
  Note: It is assumed that 64\% of the flux density
  of a pixel centred point source falls onto this pixel.
	    
\subsubsection{Method 3}
\label{sec:sem_method3}

  This method has been used at slewends, if the source was inside the
  detector aperture:
  \begin{itemize}
  \item[a)] Using the signals of all 4 pixels at the very end
            of the slew and converting them to flux densities
            assuming the source is centred on the detector. Note:
            Only 53\% of the flux of an array centred point source
            is detected, of which 21\% fall on pixel 1, 24\% on
            pixel 2, 32\% on pixel 3 and 23\% on pixel 4
            (Laureijs \cite{laureijs99}).
  \item[b)] Like case 3a, but source centred on one pixel.
  	    Note: In total 74.3\% of the source flux are seen
  	    by the 4 pixels:
  	    64\% in the source pixel, 2$\times$4.2\% in the
  	    two adjacent pixels and 1.9\% in the diagonal pixel.
  \end{itemize}
  The statistical errors are computed from the weighted
  results of the 4 pixels.

  Not \emph{all} of the ISO scientific targets are to be found in the 
  end-of-slew data: Firstly, the 4 ISO instruments view separate areas
  of the sky. 
  Slew end position (ISOPHOT) and target position (other instrument)
  can therefore differ by up to 20$^{\prime}$. Secondly, many observing
  modes, especially for ISOPHOT, started off-target for mapping purposes
  or to avoid strong detector transients. 

\section{Solar System Object Identification}
\label{sec:ext}

  \begin{figure}[bt!]
  \begin{center}
  \resizebox{!}{19cm}{\includegraphics{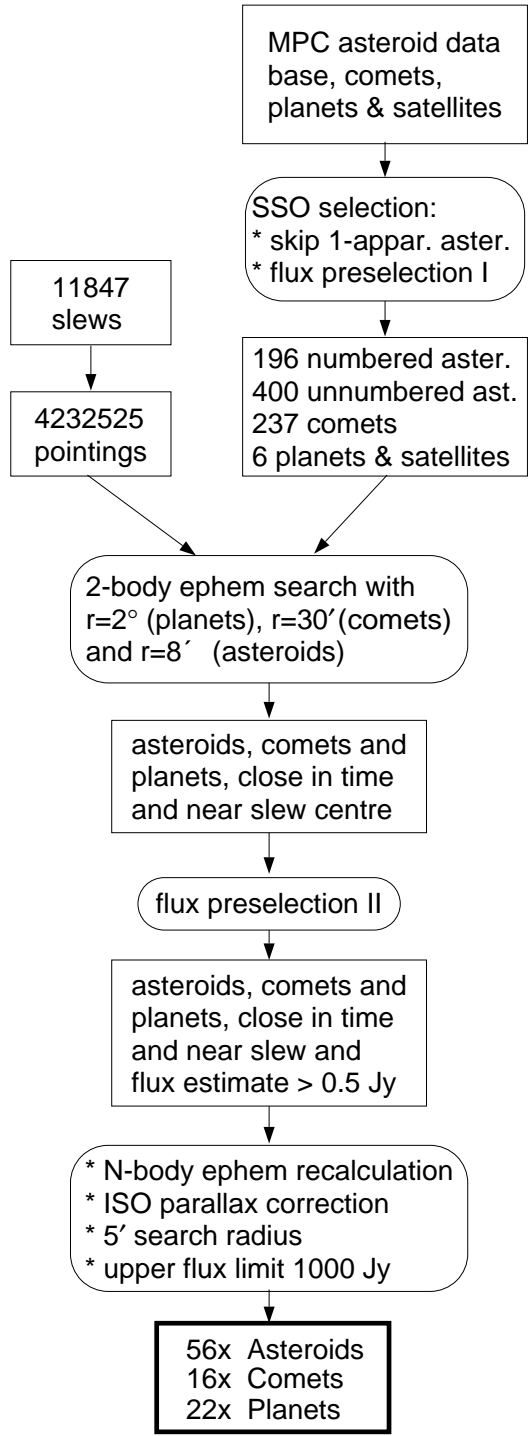}}
  \caption{Extraction procedure for ISOSS slew data
           to find SSO candidates, based only on
           pointings and timings of ISO slews and
	   expected 170\,$\mu$m fluxes. The flux preselection
	   is described in the text.
           \label{fig:ssoextraction}}
  \end{center}
  \end{figure}

  The identification and separation of moving solar system targets
  from the Serendipity slews is difficult: The Serendipity slew data
  consist of very narrow stripes across the sky, lacking, to first
  approximation, any redundancy. Additionally, the colour
  information is missing and in the far-IR region
  the cirrus confusion is a serious problem.
  Therefore, the solar system object identification was done on basis of
  accurate ephemeris calculations and model flux estimates
  (see M\"uller \cite{mueller01}).

  The Serendipity slew data consist of 11\,847 slews with a
  total length of 141\,411$^{\circ}$. The slews were
  cut in 4\,232\,525 individual pointings of
  approximately 2$^{\prime}$ length. Each of these pointings had
  to be checked against SSOs. On 20th of March 2000, the Minor Planet
  Center archive consisted of 68840 asteroids
  (14308 numbered, 24598 unnumbered with multiple-opposition orbits
  and  29934 unnumbered with single-opposition orbits) and 237 comets.
  Additionally, the outer planets and their satellites had to be
  included, leading to a total of approximately $3 \cdot 10^{11}$ ephemeris
  calculations. It was therefore necessary to preselect the number of 
  SSOs considerably and to invent fast search procedures.

\subsection{SSO Preselection}

  To facilitate and speed up the search process, only SSOs which
  at maximum are brighter than the sensitivity limit of 1\,Jy at 170\,$\mu$m
  have been considered:
  The outer planets (Mars, Jupiter, Saturn, Uranus, Neptune and Pluto)
  and their satellites were included. The inner planets were not
  visible for ISO. Due to the difficulties of predicting the brightness
  of active comets, no initial flux preselection was done for the 237 comets.
  The unnumbered single-apparition asteroids have not been considered
  because of possible large ephemeris uncertainties and generally
  too low brightness at 170\,$\mu$m (except for a few Near-Earth
  asteroids which can reach this flux limit at extremely close
  encounters).
  The preselection of numbered and unnumbered multi-apparition
  asteroids was based on conservative flux calculations, using
  a simplified Standard Thermal Model (STM, Lebofsky et al.
  \cite{lebofsky86}, \cite{lebofsky89})
  and assuming a non-rotating spherical object. The following
  conservative input values were used: $p_v = 0.02$, $G=-0.12$, $\epsilon=1.0$,
  $\eta=0.7$. The input diameter was calculated using $p_v$ and
  the absolute magnitude H with:
  $\log\,p_v = 6.259 - 2\,\log\,D_{eff} - 0.4\,H$. 
  In Flux Preselection~I, with a cut limit of 1.0\,Jy,
  the object was assumed to be located at perihelion during opposition.
  This reduced the number of asteroids by 98\% from 
  38906 to 596. In Flux preselection~II, with a cut limit of 0.5\,Jy,
  the real calculated distances (r, $\Delta$) were used.

  For the comets we determined 36 ``hits'' where the object was within
  5$^{\prime}$ from the slew centre and with the comet being within 3\,AU
  (for Hale-Bopp 5\,AU) from the Sun. A simple flux estimate lowered
  the number to 16 possible candidates. The 170\,$\mu$m
  flux estimate was based on an assumed dust albedo $A=0.10$ and a temperature
  of dust particles of $T_0=330$\,K at the heliocentric distance of $r=1$\,AU
  through the following formula:
  \begin{equation}
    F_{\nu} = B(\lambda,T) \cdot f \cdot \frac{R^2}{4\Delta^2}
  \end{equation}
  with the temperature $T=\sqrt[4]{T_0^4 \cdot (1-A)/r^2}$ and
  $\Delta$ the geocentric distance to the comet. We assumed a dense
  central coma of 10\,000\,km radius with a filling factor of $f=10^{-4}$
  and 1/R brightness profile out to a distance of 50\,000\,km.
  The corresponding model predictions fitted nicely the published results by
  Campins et al.\ \cite{campins90} for comet \object{P/Tempel 2} and
  by Hanner et al. \cite{hanner94} for comet \object{Mueller 1993a}.
  This model was also used for initial flux estimates for the preparation
  of ISO comet observations (Gr\"un, private communication).\\

\subsection{Search Radius}

  The search radius for each pointing had to be much bigger than the
  real $3^{\prime}\times3^{\prime}$ field of view of the detector
  for several reasons: 1) slew position uncertainties (up to 2$^{\prime}$,
  Stickel et al. \cite{stickel00});
  2) uncertainties of the 2-body unperturbed ephemeris, based on 200-day
  epoch orbital elements (up to a few arcmin); 3) the ISO parallax
  (up to 3$^{\prime}$ for close encounters at 0.5\,AU).
  In the first iteration, the search radii were set to 8.1$^{\prime}$ for
  asteroids (3$^{\prime}$ for ephemeris uncertainties, 3$^{\prime}$ for 
  ISO parallax, 2.1$^{\prime}$ for the centre-corner distance of
  the C200 array), to 30$^{\prime}$ for comets (to account for extended
  structures) and to 2$^{\circ}$ for the bright planets (to account for
  possible straylight influences). In the second iteration, after the
  ephemeris recalculation with an N-body programme and after parallax
  corrections, the search radius was uniformly set to 5$^{\prime}$.
  Additionally, all identified slews from the first iteration were
  marked, because of possible influences from bright SSOs.

\subsection{Search Procedure}

  Figure~\ref{fig:ssoextraction} summarizes the procedure in detail,
  giving also the input and output number of targets.
  With this procedure it was possible to reduce the initially estimated
  $10^{11}$ ephemeris calculations to $10^{9}$ 2-body and $10^{4}$ 
  N-body calculations.
  The final potential hits of 56 asteroids, 16 comets and 22 planets
  fulfilled the flux requirements at the actual time of the observation
  and were located within 5$^{\prime}$ of the slew. 
  Note that we count each encounter of a slew with an SSO as
  ``hit''. The actual numbers of different objects in this list are
  21 asteroids, 7 comets and 2 planets.
  These results, based on pure pointing and timing information, are strongly
  influenced and biased by the unequal distribution of the slews in the sky,
  satellite visibility constraints and the ISO observing programme itself.
  The relative large number of comets is
  due to the weak flux limit and it was clear that not all of them would be
  bright enough to be detected.

\section{Calibration Results}
\label{sec:calres}

  ISOSS observations of Uranus, Neptune and well known bright
  asteroids enabled us to improve and extend the existing ISOSS
  calibration (Method~1). They also allowed us to estabish the calibration
  of new source and flux extraction methods, namely
  Methods~2, 3a and 3b (see Sect.~\ref{sec:obs}).
  
  The Uranus and Neptune models are based on Griffin \& Orton
  (\cite{griffin93})
  and Orton \& Burgdorf (priv. comm.), respectively.
  For Ceres, Pallas, Juno and Vesta a thermophysical model (TPM)
  (Lagerros \cite{lagerros96}; \cite{lagerros97}; \cite{lagerros98})
  was used to predict their brightnesses at the times of the observations.
  The TPM and its input parameters are described
  in M\"uller \& Lagerros (\cite{mueller98}) and in M\"uller et al.\
  (\cite{mueller99}). The quality and final accuracy of TPM predictions
  are discussed in M\"uller \& Lagerros (\cite{mueller02a}). The general
  aspects of asteroids as calibration standards for IR projects
  are summarized in M\"uller \& Lagerros (\cite{mueller02b}).
  
  Photometric measurements of different astronomical sources can
  be compared on bases of colour corrected monochromatic fluxes
  at a certain wavelength or on basis of band pass fluxes.
  In this calibration section all model fluxes have been modified by an
  ``inverse colour correction''
  in a way that they correspond to ISOSS band pass measurements
  of a constant energy spectrum ($\nu F_{\nu} = const.$). This
  implied inverse colour correction terms of 1.09$^{-1}$ for Uranus
  and Neptune (both have temperatures at around 60\,K at 170\,$\mu$m) 
  and 1.17$^{-1}$ for the bright main-belt asteroids (assumed far-IR
  temperature of 180-200\,K), see also the colour correction tables
  in ``The ISO Handbook, Volume V'', Laureijs et al. (\cite{laureijs00}).

\subsection{Method 1\label{sct:method1}}
\label{sec:calres_method1}

  \begin{table*}[bt!]
  \begin{center}
  \begin{tabular}{rrllll}
  \hline
  \noalign{\smallskip}
  TDT & Date/Time & SSO & \fobs       & \fmodel      & \fobs/\fmodel \\
  No. &           &     & (Jy)        & (Jy)         &               \\
  (1) & (2)       & (3) & (4)         & (5)          & (6)           \\
  \noalign{\smallskip}
  \hline
  \noalign{\smallskip}
  07881200 & 03-FEB-96 09:46:42 & (4) Vesta  &  28.9 &  39.6 & 0.73 \\
  10180400 & 26-FEB-96 06:20:00 & (4) Vesta  &  30.7 &  54.2 & 0.57 \\
  14080700 & 05-APR-96 15:39:10 & Neptune    & 145.1 & 271.3 & 0.53 \\
  23080100 & 03-JUL-96 20:15:13 & (2) Pallas &  15.2 &  27.5 & 0.55 \\
  32181100 & 03-OCT-96 04:38:13 & Neptune    & 153.2 & 279.8 & 0.55 \\
  42283300 & 11-JAN-97 22:47:14 & (3) Juno   &  12.6 &  12.0 & 1.05 \\
  34480700 & 26-OCT-96 00:25:16 & Neptune    & 159.1 & 272.7 & 0.58 \\
  69880600 & 13-OCT-97 21:27:19 & Neptune    & 160.7 & 277.6 & 0.58 \\
  70681100 & 22-OCT-97 02:44:35 & Neptune    & 158.0 & 274.9 & 0.57 \\
  71381000 & 29-OCT-97 05:06:45 & Neptune    & 168.3 & 272.7 & 0.62 \\
  71980500 & 03-NOV-97 22:46:18 & Neptune    & 149.9 & 271.3 & 0.55 \\
  72081500 & 05-NOV-97 01:19:05 & Uranus     & 395.7 & 672.8 & 0.59 \\
  72081600 & 05-NOV-97 01:57:38 & Neptune    & 156.1 & 270.5 & 0.58 \\
  76280400 & 16-DEC-97 13:22:05 & (1) Ceres  &  31.5 &  52.4 & 0.60 \\
  79781500 & 21-JAN-98 00:30:12 & (4) Vesta  &  24.2 &  23.9 & 1.01 \\
  \noalign{\smallskip}
  \hline
  \end{tabular}
  \caption{Results for Method~1.
           The model fluxes are multiplied by 1.09 (planets) and
           by 1.17 (asteroids) to account for the spectral shape
           differences between $\nu F_{\nu} = \mathrm{const.}$ (assumed 
           spectrum in the ISO calibration) and the real object
           spectrum. Column~(4) contains the
	   FCS calibrated fluxes.
           \label{tbl:method1}}
  \end{center}
  \end{table*}

\begin{figure}[bt!]
  \begin{center}
    \rotatebox{90}{\resizebox{!}{9.0cm}{\includegraphics{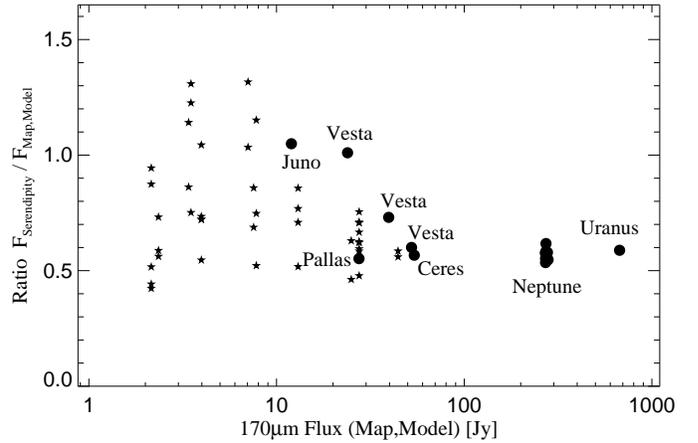}}}
  \caption{The ratio of Serendipity slew flux densities and
           model predictions for reliable Uranus, Neptune, Ceres,
           Pallas, Juno and Vesta observations. For bright sources,
           the Serendipity slews miss some flux.
           \label{fig:cal}}
  \end{center}
  \end{figure}

  ISOSS crossings over planets and asteroids, which were detected by the 
  Automatic Point Source Extractor (Stickel et al.\ 2000),
  are listed in Table~\ref{tbl:method1},
  where the columns are: (1) TDT number of
  the slew, (2) date and Universal Time at the moment of the SSO
  observation, (3) name  of the solar system  object,
  (4) observed flux density, (5) predicted flux density, (6)
  ratio between observed and modeled flux density (see also
  Fig.~\ref{fig:cal}). The ISOSS results are
  the FCS calibrated band fluxes. The model predictions were modified
  by an inverse colour correction to make them comparable with the
  ISOSS measurements (see above).
  All list entries of Uranus, Neptune, Ceres, Pallas and Vesta give a ratio
  between observed and model flux of (0.58 $\pm$ 0.05), for fluxes
  larger than about 25\,Jy. At fluxes below 25\,Jy (only 2 cases) the
  ISOSS to model ratios are close to 1.0.
  This is in excellent agreement with the results of Stickel et al.\
  (\cite{stickel00}).
  They showed that ISOSS slew fluxes of 12 selected galaxies were
  systematically lower than fluxes derived from dedicated maps. To
  bring the fluxes from mapping and slewing into agreement
  ISOSS fluxes larger than $\approx$ 30\,Jy were corrected with an
  estimated constant scaling factor of 2, while lower fluxes were scaled
  with a flux dependent correction function.
  Table~\ref{tbl:method1} represents therefore the first direct flux
  calibration of the PHT Serendipity Mode as compared to the previously
  used indirect method of flux ratios between PHT22 raster maps and slew
  results.
  
  Figure~\ref{fig:cal} shows the ratios
  between the flux densities derived from ISOSS and the 170\,$\mu$m
  model predictions. The stars represent the results from dedicated
  calibration measurements (Stickel et al. \cite{stickel00}), the
  filled circles are values from Table~\ref{tbl:method1}.
  Uranus, Neptune, Ceres, Pallas, Juno and Vesta, serendipitously seen
  by ISOSS, provide now a reliable calibration at higher flux densities.
  
\subsection{Method 2}
\label{sec:calres_method2}

  Table~\ref{tbl:method2} summarizes the values which were
  derived from the solar system far-IR standards for 
  slow slewing speeds, saturated measurements and sources
  outside the slews. These measurements were rejected by
  the source extraction procedures of Method~1.
  The table columns are: (1--6) same as in Table~\ref{tbl:method1}, (7) slew
  speed category at the moment of the SSO observation, (8) additional
  remarks. 

  \begin{table*}[h!tb]
  \begin{center}
  \begin{tabular}{rrllllll}
  \hline
  \noalign{\smallskip}
  TDT & Date/Time & SSO & \fobs       & \fmodel      & \fobs/\fmodel & Slew speed & Remarks \\
  No. &           &     & (Jy)        & (Jy)         &               &            &         \\
  (1) & (2)       & (3) & (4)         & (5)          & (6)           & (7)        & (8)     \\
  \noalign{\smallskip}
  \hline
  \noalign{\smallskip}
09380600 & 18-FEB-96 15:10:15 & (1) Ceres &   $>$46  &  73.8 &   $>$0.62    & slow     & ok \\
29280600 & 04-SEP-96 00:31:59 & (1) Ceres &   $>$54  &  67.2 &   $>$0.80    & slow     & very high bgd. \\
32880600 & 09-OCT-96 23:11:09 & Neptune   &   $>$155 & 277.8 &   $>$0.56    & slow     & ok \\
36381700 & 14-NOV-96 04:34:23 & Neptune   & $\gg$50  & 267.2 & $\gg$0.19    & moderate & outside \\
54480800 & 13-MAY-97 12:58:55 & Uranus    & $\gg$236 & 700.1 & $\gg$0.34    & moderate & saturated \\
55280300 & 21-MAY-97 06:17:09 & Uranus    & $\gg$70  & 709.5 & $\gg$0.10    & stop     & outside \\
69880200 & 13-OCT-97 17:39:49 & Uranus    & $\gg$265 & 700.1 & $\gg$0.38    & moderate & saturated \\
69880500 & 13-OCT-97 20:48:47 & Uranus    & $\gg$245 & 700.1 & $\gg$0.35    & moderate & saturated \\
71480300 & 29-OCT-97 23:34:32 & Uranus    & $\gg$202 & 680.8 & $\gg$0.30    & moderate & saturated \\
87481000 & 07-APR-98 14:36:41 & Uranus    & $\gg$232 & 652.0 & $\gg$0.36    & moderate & saturated \\
  \noalign{\smallskip}
  \hline
  \end{tabular}
  \caption{Results for Method~2.
           The model fluxes are multiplied by 1.09 (planets) and
           by 1.17 (asteroids) to account for the spectral shape
           differences between $\nu F_{\nu} = const.$ (assumed
           spectrum in the ISO calibration) and the real object
           spectrum. The values in Col.~(4) are
           already corrected for the individual pixel point-spread
           function ($F_{psf}=0.64$).
           \label{tbl:method2}}
  \end{center}
  \end{table*}

  The results from Method~2 show that also difficult slew data with
  either saturated pixels, objects slightly outside the array or
  slow speeds can be used to derive useful lower limits for interesting
  sources. As the satellite still moves the flux loss corrections
  from Method~1 have to be applied to get the best lower
  limits. In fact, for the 2 unproblematic hits (TDT 9380600 and
  32880600) with neither saturated signals nor large impact parameters,
  the flux loss correction brings the ISOSS fluxes within 10\% of the
  model predictions.

\subsection{Method 3}
\label{sec:calres_method3}

\begin{figure}[h!tb]
  \resizebox{\hsize}{!}{\includegraphics{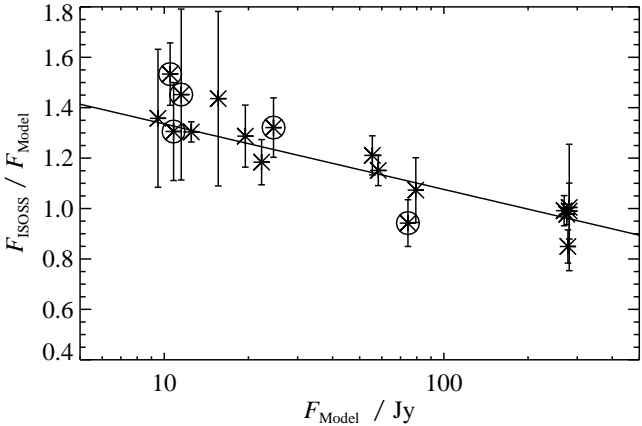}}
  \caption{The ratio of Serendipity flux densities from Method~3a and
  model predictions for Neptune, Ceres, Pallas, Juno and Vesta
  observations. 
  Error bars are statistical errors from the individual
  pixel results. Circles encompass data points from slews which had to be
  calibrated with the default calibration; in these cases, the true
  uncertainties exceed the given statistical errors.
  A flux dependency similar as in Fig.~\ref{fig:cal}
  (Method~1) can be seen. 
  \label{fig:method3}}
  \end{figure}

  At the slewend, when the satellite does not move anymore, the 
  ISOSS data can in principle be treated as normal C200 photometric
  data. Two ideal cases -- source centred on the array (Method~3a)
  and source centred on one pixel (Method~3b) -- can be distinguished.
  The results on the bright sources for both methods are summarized in
  Table~\ref{tbl:method3}, where the columns are the same as in
  Table~\ref{tbl:method1}. 
  The uncertainties in the table, given in brackets, are statistical
  errors of weighted results from all 4 pixels. The results of
  Method~3 are compared with the model predictions in
  Fig.~\ref{fig:method3}. 

  \begin{table*}[h!tb]
  \begin{center}
  \begin{tabular}{rrllll}
  \hline
  \noalign{\smallskip}
  TDT & Date/Time & SSO & \fobs       & \fmodel      & \fobs/\fmodel \\
  No. &           &     & (Jy)        & (Jy)         &               \\
  (1) & (2)       & (3) & (4)         & (5)          & (6)           \\
  \noalign{\smallskip}
  \hline
  \noalign{\smallskip}
09380500 & 18-FEB-96 07:11:04 & (1) Ceres  &  70.1(6.9)  &  74.4 & 0.94 \\
15480200 & 19-APR-96 03:42:10 & Neptune    &   269(11.1) & 275.3 & 0.98 \\
23781000 & 11-JUL-96 04:13:16 & (3) Juno   &  12.9(2.6)  &   9.5 & 1.35 \\
25180400 & 24-JUL-96 23:47:09 & (2) Pallas &  26.4(2.0)  &  22.3 & 1.18 \\
26580800 & 08-AUG-96 02:50:48 & (2) Pallas &  25.1(2.4)  &  19.5 & 1.29 \\
27580200 & 18-AUG-96 02:17:46 & (1) Ceres  &  85.2(10.2) &  79.4 & 1.07 \\
32880500 & 09-OCT-96 21:54:26 & Neptune    &   236(18.3) & 277.8 & 0.85 \\
35680200 & 06-NOV-96 20:07:37 & Neptune    &   267(16.1) & 269.3 & 0.99 \\
38781200 & 07-DEC-96 23:52:44 & (3) Juno   &  22.4(5.4)  &  15.6 & 1.44 \\
41980900 & 08-JAN-97 18:53:29 & (3) Juno   &  16.3(0.5)  &  12.5 & 1.30 \\
51080600 & 09-APR-97 10:54:42 & (2) Pallas &  14.1(2.1)  &  10.8 & 1.30 \\
51080800 & 09-APR-97 15:19:52 & (2) Pallas &  16.1(1.3)  &  10.5 & 1.53 \\
51380100 & 12-APR-97 04:33:19 & (2) Pallas &  16.7(3.9)  &  11.5 & 1.45 \\
53980100 & 08-MAY-97 03:39:36 & Neptune    &   282(70.4) & 280.8 & 1.00 \\
53980300 & 08-MAY-97 11:11:10 & Neptune    &   278(31.1) & 280.8 & 0.99 \\
54581400 & 14-MAY-97 10:49:26 & (1) Ceres  &  67.2(4.3)  &  55.5 & 1.21 \\
57581500 & 13-JUN-97 13:53:04 & (4) Vesta  &  32.5(2.9)  &  24.6 & 1.32 \\
74881000 & 03-DEC-97 02:21:54 & (1) Ceres  &  67.1(3.5)  &  58.3 & 1.15 \\
    \noalign{\smallskip}                                                 
    \noalign{\smallskip}                                                 
53880300 & 07-MAY-97 07:54:19 & (1) Ceres  &  43.8(4.9)  &  52.5 & 0.83 \\
61580800 & 23-JUL-97 02:07:07 & (4) Vesta  &  31.1(1.2)  &  34.4 & 0.91 \\
  \noalign{\smallskip}
  \hline
  \end{tabular}
  \caption{Results for Method~3 (upper part: 3a, lower part 3b).
           The model fluxes are multiplied by 1.09 (planets) and
           by 1.17 (asteroids) to account for the spectral shape
           differences between $\nu F_{\nu} = \mathrm{const.}$ (assumed
           spectrum in the ISO calibration) and the real object
           spectrum. The uncertainties in the table, given in brackets, are
           statistical errors of weighted results from all 4 pixels.
           \label{tbl:method3}}
  \end{center}
  \end{table*}

  The 5 Neptune measurements (Method~3a) agree nicely with the
  model predictions (Observation/Model: 0.96$\pm$0.09).
  For the fainter asteroids the Method~3a overestimates the flux
  systematically by 10--50\%, depending on the brightness level
  (see Fig.~\ref{fig:method3}).
  The discrepancy between bright and faint sources is probably due to
  detector nonlinearities, 
  which are not corrected in the OLP\,7 Serendipity Mode
  data, and which could be responsible for the flux dependency of the
  scaling factor (see Sect.~\ref{sct:method1}). 
  A comparison of Fig.~\ref{fig:method3} with Fig.~\ref{fig:cal} supports this
  explanation, as both diagrams show a decrease in the detector
  signals for bright sources. The fast slewing on the other hand,
  which affects Method~1 but not Method~3, could be responsible for
  the generally too low ISOSS fluxes in Fig.~\ref{fig:cal}.
 
  Both options of Methods~3 open a powerful new possibility to
  evaluate the 170\,$\mu$m fluxes of many scientific ISO targets,
  which are quite often covered in the end of slews before the
  intended science programme started.
  
\subsection{Pointing Comparison}
\label{sec:calres_pointing}

  The N-body ephemeris calculations for our SSOs included
  a transformation from geocentric to ISOcentric frame.
  The maximal geo-/ISOcentric parallax corrections 
  were: 737.7$^{\prime \prime}$ for the Apollo asteroid (7822)~1991~CS,
  336.6$^{\prime \prime}$ comet P/Encke and 61.2$^{\prime \prime}$ for
  Mars. The final accuracy of the ISOcentric SSO ephemeris has been
  estimated to about 1--2$^{\prime \prime}$.
  
  The ISOSS signal pattern, i.e.\ the relative signals of the 4 pixels,
  is a very sensitive indicator of the exact position of
  the source within the detector array.
  All close encounters have been checked by eye for
  discrepancies between predicted slew offsets and the signal patterns.
  No disagreement was found, which implies that the predicted SSO positions
  and the slew positions agree with each other within 30$^{\prime \prime}$,
  corresponding to 1/3 pixel width. In case of non-detections, the SSOs
  were either too faint, or they were actually just
  outside the slew. This high pointing accuracy allowed
  us to give upper limits (depending on the background) in cases
  when the source was crossed by the slew but no signal was
  detected (see also Sect.~\ref{sec:sem_method2}).
  In slew direction the position accuracy is better
  than 1$^{\prime}$, limited by fast slewing in combination with
  the detector read-out frequency.
  
\section{Scientific Results}
\label{sec:scires}

  The procedure from Sect.~\ref{sec:ext} resulted in a list of
  potential SSO candidates in the ISOSS. Mainly for the following
  reasons, not all of them were visible in the slew data:
  \begin{itemize}
  \item[1.)] The structured and bright cirrus background caused
	     source confusion and limited the point source extraction
	     \footnote{Even bright sources like \object{Ceres}
	               are sometimes difficult to analyse if they
		       are in regions of high background like at
		       $\lambda_{ecl.}=90^{\circ}$ and
		       $\lambda_{ecl.}=270^{\circ}$, where the ecliptic
		       crosses the galactic plane.}.
	     It also affected strongly the analysis of extended structures,
	     like from a cometary coma.
  \item[2.)] The sensitivity limit of approximately 1\,Jy at 170\,$\mu$m
             allowed only the detection of bright asteroids and comets,
	     which are already well known through other observing programmes
	     and techniques (IRAS, occultation measurements, radar, ...)
	     and through dedicated ISO measurements.
  \item[3.)] The source extraction from a 4-pixel camera is difficult due
             to the high slewing speed and the variety of impact parameters.
	     The resulting fluxes or flux limits had usually larger error bars
             than comparable pointed observations, where sources were usually
	     either centred in the C200 array or on a single pixel.
  \item[4.)] For some faint sources the allowed maximal offset of 5\arcmin\
             was too large to produce a noticable signal increase.
  \end{itemize}

  In some cases the object was visible, but a reliable flux determination
  from the ISOSS was not possible. In these cases upper or lower limits
  are given.

  Tables \ref{tbl:obsres-a1}, \ref{tbl:obsres-a2} and \ref{tbl:comets_geometry}
  include all SSO predictions which are within 5$^{\prime}$ of
  the slew centre, fulfill the flux requirements and have not been used
  in Sect.~\ref{sec:calres}.
  For the scientific comparison between ISOSS fluxes and model predictions
  we did the following calibration steps: We determined the ISOSS calibrated
  inband fluxes through the different methods and corrected them by an
  estimated factor based on the slopes visible in Figs.~\ref{fig:cal} and
  \ref{fig:method3}. As a last step we applied the colour correction to
  obtain monochromatic flux densities at 170\,$\mu$m
  (ISOSS values in Tables~\ref{tbl:obsres-a1}, \ref{tbl:obsres-a2}).
  
\subsection{Planets}
\label{sec:scires_planets}

  The inner planets were not visible for ISO due to pointing
  constraints. Mars, Jupiter and Saturn exceeded the saturation
  limits, Pluto was below the 1.0\,Jy limit. Therefore, only
  Uranus and Neptune were seen, but already used to extend the ISOSS
  calibration (Sects.~\ref{sec:calres_method1}, \ref{sec:calres_method2}
  and \ref{sec:calres_method3}). But the bright planets had to be included
  in the search programme with the objective to identify close-by slews.
  For the extremely IR bright planets, diffraction effects of the optical
  system in combination with certain satellite-planet constellations produced
  bright spots, spikes and ring like structures around the planets,
  which are visible in the slew data. Mars, Jupiter and Saturn, with
  170\,$\mu$m brightnesses between 10\,000 and 400\,000\,Jy, influenced
  slews up to 1$^{\circ}$ distance, Uranus and Neptune ($>$ 200\,Jy)
  up to 10$^{\prime}$ distance. These slews have been identified
  (with the above mentioned SSO extraction method in combination with a large
  search radius) and the planet influence can now be taken into
  account for further scientific catalogues based on ISOSS. 
  Some planetary satellites are bright enough to be visible in principle.
  However, close to Jupiter (the maximal distance for the Galilean
  satellites is about 11$^{\prime}$) and Saturn (the maximal angular
  distance for the 8 largest satellites is about 10$^{\prime}$) no
  170\,$\mu$m fluxes can be derived, due to the strong planet influences.
  
  The measured \object{Uranus} and \object{Neptune} flux values from
  Tables~\ref{tbl:method1}, \ref{tbl:method2} and \ref{tbl:method3}
  can also be used as input to future models of planetary atmospheres.
  Current models are based on Voyager IRIS data from 25 to 50\,$\mu$m
  and sub-millimetre data beyond 350\,$\mu$m (Griffin \& Orton
  \cite{griffin93}) with an interpolation in between. The ISOSS
  observations make this wavelength range directly accessible for
  model tests in the far-IR.

\subsection{Asteroids}
\label{sec:scires_asteroids}

  After establishing new methods for the flux calibration, based now
  additionally on measurements of \object{Uranus}, \object{Neptune},
  \object{Ceres}, \object{Pallas}, \object{Juno} and \object{Vesta},
  the monochromatic flux densities at 170\,$\mu$m of the remaining
  asteroids were derived. 23 out of the 56 asteroid ``hits'' 
  were already used in this calibration context in the previous
  Sect.~\ref{sec:calres}.
  The remaining 33~hits 
  can be split in 3~groups:
  
\subsubsection{IRAS and Poor ISOSS Detection}

  18 asteroid predictions have reliable IRAS detections,
  but only low quality ISOSS detections. The reasons for the poor ISOSS
  fluxes are manifold: slew offsets, bright backgrounds,
  technical problems, low fluxes, etc.
  For these 18 asteroids Tedesco et al.\ (\cite{tedesco92})
  calculated already diameters and albedos, based on IRAS
  observations. Upper flux limits from ISOSS would therefore
  not give any new information.
	
\subsubsection{IRAS and Good ISOSS Detection}
  
  \begin{table*}[h!tb]
  \begin{center}
  \begin{tabular}{rrllllll}
  \hline
  \noalign{\smallskip}
  TDT      & Date/Time          & SSO              & \fobs  & \fmodel & Method & Slew   & Remarks \\
  No.      &                    &                  & (Jy)   & (Jy)&	& speed   &	    \\
  (1)      & (2)                & (3)              & (4)    & (5) & (6) & (7)	  & (8)     \\
  \noalign{\smallskip}
  \hline
  \noalign{\smallskip}
  11080300 & 06-MAR-96 11:42:00 & (5)~Astraea      & $<$4 & 1.3 & 2   &  fast	& upper limit, cirrus bgd.\\
  12780300 & 23-MAR-96 02:50:17 & (5)~Astraea      & $<$4 & 1.6 & 2   &  fast	& upper limit, cirrus bgd.\\
  85480200 & 18-MAR-98 13:48:12 & (7)~Iris         & $<$3 & 2.1 & 2   &  fast	& in cirrus knot \\
  25381400 & 27-JUL-96 03:22:14 & (15)~Eunomia     & 2.9  & 2.5 & 1   &  fast	& ok \\
  85480300 & 18-MAR-98 17:22:43 & (89)~Julia       & $<$3 & 1.7 & 2   &  fast	& upper limit \\
  18780100 & 22-MAY-96 05:40:00 & (344)~Desiderata &  4.5 & 5.8 & 1   &  fast	& ok \\
  18280800 & 17-MAY-96 03:50:00 & (532)~Herculina  & 5.3(1.0) & 5.5 & 3a  &  stop   & ok \\
  21780900 & 21-JUN-96 06:01:48 & (532)~Herculina  & 3.4(0.3) & 3.8 & 3a  &  stop   & ok \\
  83380500 & 25-FEB-98 11:36:26 & (1036)~Ganymed   & $<$4 & 0.1 & 2   &  fast   & upper limit \\
  \noalign{\smallskip}
  \hline
  \end{tabular}
  \caption{ISOSS observational results for asteroids with IRAS
           detections. Note, that the ISOSS results were first flux
           corrected according to Sect.~\ref{sec:calres} and then colour
           corrected.The uncertainties in the table, given in brackets for
           method 3a, are statistical errors of weighted results from
           all 4 pixels.
           \label{tbl:obsres-a1}}
  \end{center}
  \end{table*}

  7 asteroids (9 hits) have reliable IRAS
  detections and also good quality ISOSS detections (see
  Table~\ref{tbl:obsres-a1}).
  For these asteroids we could derive successfully 
  fluxes and upper limits with our newly established
  calibration, based on clear detections. The IRAS diameter
  and albedo values in the following are all taken from the
  Minor Planet Survey (MPS, Tedesco et al. \cite{tedesco92}).

  \paragraph{\object{Astraea: }}
  The main-belt asteroid \object{Astraea} has been well observed by
  different techniques, including IRAS, radar and lightcurve
  observation. The combination of all measurements led to 
  the description of the object as a rotating ellipsoid with a well
  determined spin vector (Erikson \cite{erikson00}) and axis dimensions of
  $143 (\pm 12\,\%) \times 115 \times 100$\,km 
  (Magri et al. \cite{magri99}).
  Using the TPM with default thermal parameters for main-belt
  asteroids (M\"uller et al. \cite{mueller99}) together with the
  shape, size and spin vector information gave fluxes of
  1.3 $\pm$ 0.3\,Jy and 1.6 $\pm$ 0.4\,Jy at the 2 ISOSS epochs
  (see Table~\ref{tbl:obsres-a1}). The measured ISOSS upper limits
  are in agreement with the TPM predictions.

  \paragraph{\object{Iris: }}
  Like for \object{Astraea}, a shape model has been established
  for \object{Iris} based on a combination of radiometric, lightcurve
  and occultation data (Magri et al. \cite{magri99}).
  The corresponding TPM prediction gives 2.1 $\pm$ 0.2\,Jy at
  the ISOSS epoch, with an additional lightcurve variation
  of about 25\,\% (min to max). The measured lower flux limit is
  in agreement with the calculations.

  \paragraph{\object{Eunomia: }}
  \object{Eunomia} was one of the best observed asteroids by IRAS:
  7 epochs distributed over almost one month, each time observed with
  high S/N in all 4 bands. The MPS diameter
  is 255.3 $\pm$ 15\,km and the albedo 0.21 $\pm$ 0.03. Two single
  chord occultation measurements led to diameters of $>$309 $\pm$ 5\,km
  (Overbeek \cite{overbeek82}) and $>$232\,km (Stamm \cite{stamm91}).
  The TPM prediction (based on MPS
  diameter and albedo and on shape and spin vector by Erikson
  \cite{erikson00}) gives 2.5 $\pm$ 0.5\,Jy at the ISOSS epoch,
  with the main error contribution coming from the large lightcurve
  variation. The measured ISOSS flux of 2.9\,Jy agrees within the
  errorbars.

  \paragraph{\object{Julia: }}
  IRAS observed this asteroid 4 times within 2 weeks, each time
  with high S/N in all 4 bands. The MPS diameter
  is 151 $\pm$ 3.1\,km and the albedo 0.18 $\pm$ 0.01. No shape
  or spin vector is available, but the possible lightcurve
  amplitudes range between 0.10 and 0.25\,mag (Lagerkvist et al.
  \cite{lagerkvist89}). The TPM prediction (based on MPS
  diameter and albedo together with a spherical shape) gives 1.7 $\pm$ 0.1\,Jy
  at the ISOSS epoch, with an additional maximal lightcurve variation
  of approximately 25\,\% (min to max). The measured upper limit
  agrees with this prediction.

  \paragraph{\object{Desiderata: }}
  This asteroid was observed by IRAS extensively at 9 epochs during
  a period of 2 months with high S/N in all bands. The MPS diameter
  is given with 132.3 $\pm$ 5.5\,km and the albedo 0.06 $\pm$ 0.01. 
  No shape and spin vector exists currently for \object{Desiderata},
  but a 0.17\,mag lightcurve amplitude has been stated by Lagerkvist
  et al. (\cite{lagerkvist89}). The TPM prediction (based on MPS
  diameter and albedo together with a spherical shape) gives 5.8 $\pm$ 0.5\,Jy
  at the ISOSS epoch, with an additional maximal lightcurve variation
  of approximately 17\,\% (min to max). Assuming an ISOSS observation
  at lightcurve minimum and a diameter at the lower end of MPS diameter
  range would result in a TPM flux which is only a few percent above the
  measured ISOSS value, but well within the ISOSS measurement error bars.

  \paragraph{\object{Herculina: }}
  7 IRAS observations with high S/N in either 3 or 4 bands have been
  obtained between March and October 1983. The The MPS diameter
  is given with 222.2 $\pm$ 7.6\,km and the albedo 0.17 $\pm$ 0.01. 
  The occultation diameter of 217 $\pm$ 15\,km is based on several
  chords in combination with information on the pole orientation and a
  lightcurve fit (Bowell et al. \cite{bowell78}).
  MPS and occultation diameters agree nicely. The
  complete shape and spin vector solutions derived from lightcurve
  observations are given in Erikson (\cite{erikson00}). Using the
  full information for \object{Herculina} (see also M\"uller \&
  Lagerros \cite{mueller98} for details) led to
  170\,$\mu$m fluxes of 5.5 $\pm$ 0.4\,Jy and 3.8 $\pm$ 0.3\,Jy.
  These values have
  been calculated using the exact lightcurve phase and amplitude
  at the time of the ISOSS observations. The almost perfect 
  agreement between predicted and measured fluxes (see
  Table~\ref{tbl:obsres-a1}) confirms
  in an independent way the reliable calibration of this
  new flux extraction method for the ISOSS.

  \paragraph{\object{Ganymed: }}
  IRAS saw \object{Ganymed} only twice and in both cases only
  the 25\,$\mu$m flux was useable for the radiometric
  calculations, resulting in a diameter of 31.7 $\pm$ 2.8\,km
  and an albedo of 0.29 $\pm$ 0.06. A single chord occultation
  measurement gave a lower diameter limit of 16\,km (Langans
  1985\footnote{\tt http://sorry.vse.cz/\~{}ludek/mp/world/mpocc1.txt}).
  The large lightcurve amplitude
  of 0.45\,mag (Lagerkvist et al. \cite{lagerkvist89}) adds
  more uncertainties to the model calculations. Purely based
  on the given diameter and albedo, the TPM predicts
  approximately 0.1\,Jy for the time of the ISOSS observation,
  which is well below the detection limit of this observing mode.

\subsubsection{No IRAS Detection}
  
  \begin{table*}[h!tb]
  \begin{center}
  \begin{tabular}{rrlllllll}
  \hline
  \noalign{\smallskip}
  TDT      & Date/Time          & SSO            & \fobs & \fmodel & Method & Slew & Real source & Remarks \\
  No.      &                    &                & (Jy)   & (Jy)&     & speed & offset$^\star$   &	   \\
  (1)      & (2)                & (3)            & (4)    & (5) & (6) & (7)   & (8)		& (9)	  \\
  \noalign{\smallskip}
  \hline
  \noalign{\smallskip}
  83081700 & 22-FEB-98 19:07:18 & (9)~Metis       &  3.8 & 4.1 & 1   &  fast & 0\arcmin   & ok \\
  84281000 & 06-MAR-98 16:31:55 & (9)~Metis       & $>$1 & 3.4 & 2   &  fast & 0.3\arcmin & $\sim3\sigma$ detection\\
  63681300 & 13-AUG-97 10:42:38 & (27)~Euterpe    & ---  & 1.5 & --- &  fast & 1.9\arcmin & no detection\\
  25880900 & 01-AUG-96 07:42:48 & (391)~Ingeborg  & ---  & 0.2 & --- &  fast & 2.5\arcmin & no detection\\
  71682300 & 01-NOV-97 06:03:56 & (3753)~Cruithne & ---  & 0.1 & --- &  fast & 1.0\arcmin & no detection\\
  27482100 & 17-AUG-96 01:22:01 & (7822)~1991 CS  & ---  & 0.1 & --- &  fast & 2.5\arcmin & no detection\\
  \noalign{\smallskip}
  \hline
  \end{tabular}
  \caption{ISOSS observational results for asteroids without
  	   IRAS detection. $^\star$~Closest approach to edge of closest pixel.
  	       \label{tbl:obsres-a2}}
  \end{center}
  \end{table*}

  5 asteroids (6 hits) have no IRAS detection, but fulfilled
  the conservative flux requirements for the ISOSS asteroid search
  (see Table~\ref{tbl:obsres-a2}). Unfortunately 4 
  sources (\object{Euterpe}, \object{Ingeborg}, \object{Cruithne}
  and \object{1991~CS}) have only marginal ISOSS detections
  and establishing upper flux limits was difficult.
  The results of Table~\ref{tbl:obsres-a2} per Object:

  \paragraph{\object{Metis: }}
  Kristensen (\cite{kristensen84}) has determined a size of
  190 $\pm$ 19\,km for this asteroid from an occultation event.
  A second occultation a few years later gave a high quality
  173.5\,km diameter (Stamm \cite{stamm89}, Blow \cite{blow97}).
  Recent HST
  images (Storrs et al. \cite{storrs99}) revealed an elongated disk
  with a long axis of 235\,km and a short axis of 165\,km, which
  corresponds to an effective diameter of 197\,km. Given the uncertainties
  involved we adopt the occultation result which is perfectly consistent
  with both techniques (see also Lagerros et al. \cite{lagerros99}).
  The full light curve and shape information
  has been taken from Erikson (\cite{erikson00}). The TPM predictions
  gave  4.1 $\pm$ 0.8\,Jy and 3.4 $\pm$ 0.7\,Jy, respectively
  (see Table~\ref{tbl:obsres-a2}). Adopting the HST results instead
  led to about 5\,\% and 10\,\% higher fluxes.
  
  Based on the ISOSS flux of 3.8\,Jy, the TPM allowed the calculation
  of an effective projected diameter of 178\,km and an albedo
  of $p_V=0.15$ at the epoch of the ISOSS observation.
  A possible 20\,\% ISOSS flux uncertainty would correspond to
  about 10\,\% diameter uncertainty, resulting in a
  size of the rotating ellipsoid of
  $213 \times 164 \times 132$\,km with 10\,\% minimum uncertainties.
  
  Within the different uncertainties and based on the shape and spin
  vector solutions, the results agree nicely. The 3 methods --occultation,
  HST direct imaging and ISOSS radiometric method-- led to
  comparable diameters and albedos.

  \paragraph{\object{Euterpe: }}
  No IRAS observations are available. We used instead the largest 
  extension from an occultation measurement (Dunham \cite{dunham98})
  together with a shape and spin-vector model (Erikson \cite{erikson00}),
  H, G values (Piironen et al. \cite{piironen97}) and an albedo of
  0.13 related to the occultation cross section. The TPM prediction
  was 1.5\,Jy with a large uncertainty due to the limited size knowledge. 
  This is well within the detection limits, but the source was too far
  from the slew center to determine an upper flux limit.

  \paragraph{\object{Ingeborg: }}
  There exists hardly any information about this asteroid. Based on its
  H-value of 10.1\,mag, together with a typical S-type (Tholen \cite{tholen89})
  albedo of 0.155 one can calculate an approximate diameter of 32\,km.
  The corresponding flux calculation for the ISOSS epoch gave 0.2\,Jy,
  which is clearly below the detection limit. Even under the assumption
  of an extreme albedo of 0.03 the asteroid flux at 170\,$\mu$m
  would only be 1.3\,Jy and therefore hardly detectable. Like in the
  case of \object{Euterpe}, \object{Ingeborg} had a slew center offset 
  which was close to the maximal allowed 5\arcmin.

  \paragraph{\object{Cruithne: }}
  \object{Cruithne} is currently the only known object on a horseshoe
  orbit around Earth (Christou \cite{christou00}). It was also part of
  a special near-Earth object observations
  programme (Erikson et al. \cite{erikson00a}). Based on an unweighted
  mean of typical C and S-type asteroids ($p_V=0.12$) and an H-value
  of $H=15.13\pm0.05$, they calculated a diameter of 3.7\,km and
  a slow rotation period of 27.4 hours. Although the observing
  geometry with only 0.37\,AU from Earth was almost ideal, the 170\,$\mu$m
  flux was only 0.1\,Jy. Even an extremely low albedo (leading to a diameter
  of about 8\,km) would only give 0.3\,Jy well below the detection limit.
  Therefore an upper limit from a background analysis would not give
  any new information.

  \paragraph{\object{1991 CS: }}
  The case of \object{1991 CS} is similar to \object{Cruithne}: A near-Earth
  asteroid, at only 0.14\,AU from Earth at the time of the ISOSS slew and
  with an H-value of 17.4\,mag. A radar campagne resulted in an estimated
  diameter of 1.1\,km, an albedo of 0.14 and a rotation period of 2.39\,hours
  (Pravec et al. \cite{pravec98}). The TPM predicts a 170\,$\mu$m
  flux below 0.1\,Jy and even for extreme albedo values the flux would be below
  0.3\,Jy and therefore not detectable for ISOSS.
 
\subsubsection{Additional Results}

  The \object{Juno} observations in Table~\ref{tbl:method1} and in
  Table~\ref{tbl:method3} have flux ratios systematically higher than
  ratios from comparable sources.
  Calibrating the ISOSS values with the corresponding newly established
  methods~1 and 3a resulted in an average observation over model ratio
  of 1.14, indicating that the model diameter of \object{Juno} is about
  7\,\% too low. M\"uller \& Lagerros (\cite{mueller02a}) analysed 11
  independent ISO observations, taken with the long wavelengths ISOPHOT
  detectors. They find a similar mean ratio of 1.13$\pm$0.10, which
  confirms the tendency to higher diameter values. Both investigations
  indicate that the effective diameter should be close to 260\,km, compared
  to the published values of 241.4\,km (M\"uller \& Lagerros \cite{mueller98})
  and 233.9$\pm$11.2\,km (Tedesco et al. \cite{tedesco92}).
  
  For \object{Vesta} the situation is not that clear. The values in
  Table~\ref{tbl:method1} are not conclusive since the \object{Vesta}
  fluxes cover the difficult transition region between little flux loss and
  the more than 40\,\% flux loss for sources brighter than 25\,Jy
  (see Fig.~\ref{fig:cal}). 
  It seems that two of the measured fluxes (TDT 07881200, 79781500)
  are higher than one would expect from other sources with
  similar brightness. This contradicts the findings by Redman et al.
  (\cite{redman98}; \cite{redman92}). They state for \object{Vesta}
  an extremely low emissivity of 0.6 in the submillimetre. Assuming
  that the emissivity is already lower in the far-IR
  would mean that the \object{Vesta} points in
  Fig.~\ref{fig:cal} should lie below the general trend and not
  above. The measurement from methods~3a (TDT 57581500) and 3b (TDT 61580800)
  agree within the errorbars with the model predictions.
  As in M\"uller \& Lagerros (\cite{mueller02a}), we see no clear
  indications of far-IR emissivities lower than the default values
  given in M\"uller \& Lagerros (\cite{mueller98})
 
\subsection{Comets}
\label{sec:scires_comets}

\subsubsection{Observational Results}
  
  The results of the positional search through the ISOSS pointing data,
  combined with the flux estimates are given in
  Table~\ref{tbl:comets_geometry}.
  The table columns are: (1--3) same as in Table~\ref{tbl:method1}, 
  (4--5) ISO-centric coordinates (2000.0), (6--7) Sun and Earth distance
  at the time of the observation, (8--9) ISOSS and model flux.
  \begin{table*}[bt!]
  \begin{center}
  \begin{tabular}{rrlllllll}
  \hline
  \noalign{\smallskip}
  TDT      & Date/Time          & SSO            & R.A.       & Dec.        & $r$   & $\Delta$ & \fobs & \fmodel \\
  No.      &                    &                & (hms)      & (dms)       & (AU)  & (AU)     & (Jy)  & (Jy)    \\
  (1)      & (2)                & (3)            & (4)        & (5)         & (6)   & (7)      & (8)   & (9)     \\
  \noalign{\smallskip}
  \hline
  \noalign{\smallskip}
  60780100 & 14-JUL-97 22:57:49 & 2P/Encke       & 14 56 28.6 & $-$63 36 09 & 1.164 & 0.264    & 5-10  & $>$50 \\
  \noalign{\smallskip}
  34881300 & 29-OCT-96 22:27:01 & 22P/Kopff      & 21 23 56.6 & $-$19 53 22 & 1.961 & 1.523    & 0.5-1 & $<$1 \\
  \noalign{\smallskip}
  23380800 & 07-JUL-96 11:31:51 & 96P/Machholz 1 & 23 10 27.5 & $-$68 19 46 & 2.052 & 1.328    & no det. & $\approx$1 \\
  \noalign{\smallskip}
  77780200 & 31-DEC-97 16:38:09 & 103P/Hartley 2 & 23 27 54.1 & $-$07 29 09 & 1.041 & 0.825    & poor det. & $\approx$25 \\
  \noalign{\smallskip}
  80280100 & 25-JAN-98 15:07:48 & 104P/Kowal 2   & 00 43 46.0 & $+$08 38 35 & 1.451 & 1.496    & 1     & $\approx$2 \\  
  \noalign{\smallskip}
  33280100 & 13-OCT-96 18:43:22 & 126P/IRAS      & 21 38 46.7 & $-$29 54 48 & 1.712 & 1.028    & ---   & $>$2 \\
  36280400 & 12-NOV-96 13:51:16 & 126P/IRAS      & 21 45 50.3 & $-$08 47 05 & 1.709 & 1.307    & 1.0   & $<$2\\
  \noalign{\smallskip}
  13481800 & 30-MAR-96 17:26:33 & C/1995 O1 $=$  & 19 42 20.5 & $-$19 43 10 & 4.867 & 5.004    & see text & \\
  16280600 & 27-APR-96 14:17:19 & Hale-Bopp      & 19 44 35.1 & $-$17 37 42 & 4.588 & 4.259    & 9.3$\pm$1.8 & \\
  31580500 & 27-SEP-96  0:05:16 & \multicolumn{1}{c}{$\prime\prime$} & 17 29 43.0 & $-$05 11 32 & 2.934 & 2.965 & 30.9$\pm$7.3 & \\
  32081300 & 01-OCT-96 17:21:44 & \multicolumn{1}{c}{$\prime\prime$} & 17 29 42.8 & $-$04 57 58 & 2.878 & 2.987 & see text & \\
  32280200 & 03-OCT-96 15:16:59 & \multicolumn{1}{c}{$\prime\prime$} & 17 29 50.7 & $-$04 52 28 & 2.856 & 2.995 & see text & \\
  32580600 & 06-OCT-96 23:28:20 & \multicolumn{1}{c}{$\prime\prime$} & 17 30 13.8 & $-$04 42 47 & 2.816 & 3.009 & see text & \\
  77081500 & 25-DEC-97  0:18:51 & \multicolumn{1}{c}{$\prime\prime$} & 06 32 55.7 & $-$64 09 08 & 3.851 & 3.683 & 43.8$\pm$4.0 & \\
  86880300 & 01-APR-98 14:02:15 & \multicolumn{1}{c}{$\prime\prime$} & 05 02 13.8 & $-$53 09 12 & 4.855 & 4.945 & see text & \\
  87380400 & 06-APR-98 16:12:29 & \multicolumn{1}{c}{$\prime\prime$} & 05 05 07.0 & $-$52 34 21 & 4.905 & 5.009 & 15.0$\pm$2.9 & \\
  \noalign{\smallskip}
  \hline
  \end{tabular}
  \caption{Observational geometry for the comets. No model values have been
           calculated for \object{C/1995 O1} (Hale-Bopp). All hits are discussed in the text.
           One predicted hit was a ``no detection'', one a ``poor detection'' and
           in one case (---) the slew length was too short.
           \label{tbl:comets_geometry}}
  \end{center}
  \end{table*}

  \paragraph{\object{2P} (Encke): }
  The comet has been detected at a slew end position on an
  extremely high background close to the galactic plane.
  The flux increase towards the comet nucleus corresponds
  to about 5--10\,Jy. The coma extension and its brightness
  profile could not be determined due to the high background
  brightness. The model flux at this close encounter with
  Earth (0.26\,AU) may have been strongly overestimated due to
  the large apparent size of the central coma which was
  assumed to be of constant brightness.
  
  \paragraph{\object{22P} (Kopff): }
  The ISOSS slew ends again on the comet, but this time
  the source is located on a clean low background. The signal
  pattern is similar to that of a point-source with
  $\sim$\,0.5--1\,Jy, 
  which is close to the detection limit. An upper flux
  limit of 2\,Jy can be given, which is in good agreement with
  the simple model calculations (Table~\ref{tbl:comets_geometry}).
  
  \paragraph{\object{96P} (Machholz 1): }
  The comet has not been detected. The position calculation
  showed that the source was just outside the slew path, but
  within the specified 5\arcmin\ search limit. The low model
  flux indicated already the difficulty to detect the coma
  or the comet nucleus. 
  
  \paragraph{\object{103P} (Hartley 2): }
  Only a poor detection of an extended source was found, although
  the comet was on a low background. ISOPHOT observations close to the
  ISOSS observing epoch show that \object{Hartley~2} had a colour
  temperature of 285\,K (Colangeli et al. \cite{colangeli99}).
  This is 30\,K colder than the calculated model temperature.
  A second reason for the discrepancy between a low ISOSS flux
  and the predicted 25\,Jy is probably the too large apparent size of
  the central coma which was assumed to be of constant brightness.
  At an Earth distance of 0.82\,AU the model comet core covers a
  significant part of the aperture. Both effects together might
  explain the model value.
   
  \paragraph{\object{104P} (Kowal 2): }
  A source of approximately 1\,Jy was detected by one pixel
  at the predicted position of the comet, but confusion with
  a close IRAS source could not be excluded.
  
  \paragraph{\object{126P} (IRAS): }
  In the first case the slew length was only 1\arcmin\ which was not
  sufficient for the data analysis. The second case was a clear
  detection by one pixel (Method 2). The derived flux of 1\,Jy
  is in agreement with the calculated upper limit of 2\,Jy.

  \paragraph{\object{C/1995 O1} (Hale-Bopp): }
  {\bf 13481800}: The slew passed over the coma with the nucleus only
  30\arcsec\ outside the closest pixel. A weak signal of
  $\sim$\,2\,\mjysr\ has been detected in this pixel.
  {\bf 16280600}: The slew ended on \object{Hale-Bopp} and the integrated
  4-pixel flux was determined to 9.3$\pm$1.8\,Jy (Method 3a).
  {\bf 31580500}: Method 3a was applicable and a 170\,$\mu$m flux of
  30.9$\pm$7.3\,Jy has been derived.
  {\bf 32081300}: Slew over the comet nucleus, with one pixel
  crossing centrally, two pixels in 1\arcmin\ distance and one pixel in
  2\arcmin\ distance. The slew crossed the nucleus under an angle of
  45$^{\circ}$ relative to the orientation of the dust tail
  ($PsAng$\footnote{$PsAng$: The position angle of the
   extended Sun $\rightarrow$ target radius vector as seen in the
   observer's plane-of-sky, measured counter-clock wise from reference
   frame North Celestial Pole.}$=87.8^{\circ}$,
   $ISOSS PosAng$\footnote{$ISOSS PosAng$: The position
   angle of the ISOSS slew origination as seen in the observer's
   plane-of-sky, measured counter-clock wise from reference frame
   North Celestial Pole.}$=42.0^{\circ}$).
  The measured brightness profile clearly deviates from that of a
  point-source. The asymmetric profile is stronger towards the east,
  i.e.\ on the tail-side of the nucleus.
  Due to a nearby cirrus ridge, the dust tail extension is confirmed out
  to 2\arcmin\ only (but would be probably larger on a flat background).
  {\bf 32280200}: The slew crossed the dust tail of \object{Hale-Bopp}
  under an angle of approx.\ 30$^{\circ}$ in 8\arcmin\ distance from
  the nucleus ($PsAng=87.1^{\circ}$, $ISOSS PosAng=55.8^{\circ}$).
  A signal increase at the position of the dust tail can be seen, but
  the signal pattern is difficult to discriminate from the cirrus
  structures in the background, hence quantitatively not helpful.
  At the closest comet approach of 4\arcmin\ a second signal increase
  can be seen which coincides with the position angle of the negative
  of the target's heliocentric velocity vector
  ($PsAMV$\footnote{$PsAMV$: The position angle of the extended Sun
   $\rightarrow$ target radius vector as seen in the observer's
   plane-of-sky, measured counter-clock wise from reference frame North
   Celestial Pole.}$=151.2^{\circ}$).
  The signal increase is either related to the large cometary coma at
  a distance of only $r=2.86$\,AU from the sun or a kind of trail
  formation in the direction of $PsAMV$ similar to what Reach et al.
  (\cite{reach00}) found for comet \object{Encke}.
  {\bf 32580600}: The detectors moved centrally along the dust tail
  and cross over the comet nucleus ($PsAng=85.9^{\circ}$,
  $ISOSS PosAng=85.0^{\circ}$). The measured brightness
  profile clearly deviates from a point-source profile (see
  Figure~\ref{fig:hale-bopp}). A dust tail extension of more than
  4\arcmin\ can be seen where the satellite approaches the nucleus.
  The signal in anti-tail direction decreases more rapidly
  (see also Section~\ref{sec:hb}).
  {\bf 77081500}: Method 3a was applicable again and a 170\,$\mu$m flux
  of 43.8$\pm$4.0\,Jy was derived.
  {\bf 86880300}: The slew passes in 4\arcmin\ distance ahead of the comet
   tail under an angle of approx.\ 70$^{\circ}$ with the sun direction
   ($PsAng=113.3^{\circ}$, $ISOSS PosAng=183.5^{\circ}$).
   A signal change of 2\,\mjysr\ extended over 15\arcmin\ can clearly
   be seen. Due to the viewing geometry (the phase angle is only
   11$^{\circ}$) coma and tail are difficult to separate and the signal
   increase is most likely connected to the dust emission of the extended
   coma and tail structures of \object{Hale-Bopp}. Here, as in 32280200,
   the signal maximum coincides with the $PsAMV$ angle of $1.2^{\circ}$.
   A connection might be possible between the 170\,$\mu$m signal pattern
   and large particles forming an elongated structure behind the comet
   nucleus while it is moving away from perihelion.
  {\bf 87380400}: Method 3a was applicable again and a 170\,$\mu$m flux
  of 15.0$\pm$2.9\,Jy was derived.

\subsubsection{\object{C/1995 O1} (Hale-Bopp)}
\label{sec:hb}

  \begin{figure}[h!tb]
  \resizebox{\hsize}{!}{\includegraphics{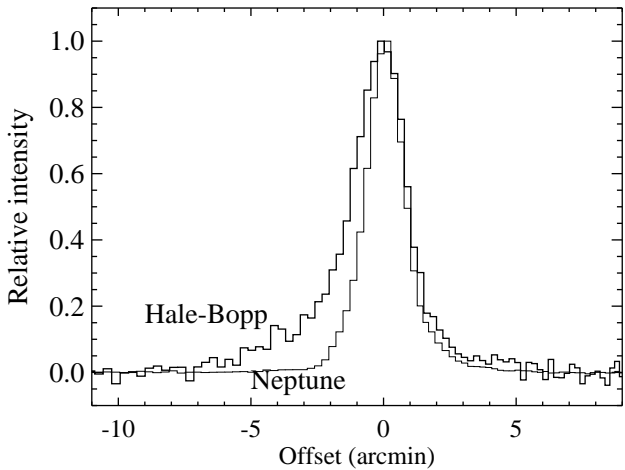}}
  \caption{An ISOSS profile of \object{Hale-Bopp} in comparison
           with the signal pattern of a point source. The slewing
           speeds for these two hits were identical and already quite low
           (2\arcmin/s), resulting in a high sampling rate. The offsets are
           defined with respect to the positions from the ephemeris 
           calculations and have not been shifted for the two sources relative
           to each other. \object{Hale-Bopp} shows an asymmetrically extended
           profile. Note that the profiles are not deconvolved.
            \label{fig:hale-bopp}}
  \end{figure}
  
  Figure~\ref{fig:hale-bopp} illustrates a measured signal profile
  from a central slew over \object{Hale-Bopp} (TDT 32580600) in
  comparison with a slew over the point source \object{Neptune}
  (TDT 72081600). For these two detections, the geometrical
  configurations and the slewing speeds (2\arcmin/s) have been
  identical. The detectors moved first centrally over the dust
  tail (left side of the peak) and then over the nucleus of
  \object{Hale-Bopp} (peak). An asymmetric signal profile between
  $-6$\arcmin\ and $+4$\arcmin\ can be seen. In case of \object{Neptune}
  the signal increase starts approximately 2\arcmin\ ahead of the
  true position, which is related to a combination of the
  Airy disk with the slew speed and read out frequency
  (see also Hotzel et al. \cite{hotzel01}). The slight shift
  between the two peaks is most probably related to the possible
  positional uncertainties of ISOSS data (see
  Section~\ref{sec:calres_pointing}). This  \object{Hale-Bopp} asymmetry
  has not been seen in dedicated 170\,$\mu$m maps (Peschke et
  al. \cite{peschke99}) which were taken at $r=3.904$\,AU (as compared
  to $r=2.816$\,AU in Figure~\ref{fig:hale-bopp}).

  The fluxes derived from Method 3a can be compared to results from
  Gr\"un et al. \cite{gruen01} through the following corrections:
  1) Flux corrections according to Figure~\ref{fig:method3};
  2) Normalization to a standard aperture diameter of 23\arcsec\,
     assuming that the coma brightness scales linearly with
     aperture diameter ($c_{a23}=0.120$);
  3) Point-spread-function correction which takes into account
     the differences of a point source and a $1/\rho$-coma
     ($c_{psf}=1.092$);
  4) Colour correction\footnote{There seems to be a wrong application
     (multiplication instead of division) of the colour correction
     factor by Gr\"un et al. (\cite{gruen01}). For consistency,
     we do however apply all corrections as given in their paper.}
     which changes for different distances
     from the sun ($c_{colour}=f(r)$).

  The first two measurements (16280600, 31580500) were obtained
  on the same days as the ones in Gr\"un et al. \cite{gruen01}.
  The calibrated and reduced 23\arcsec\ fluxes agree within
  the errorbars. The third observation (77081500), taken 5\,days
  earlier than the dedicated \object{Hale-Bopp} observation,
  lead to a  flux of 5.64$\pm$0.56\,Jy, as compared to 2.66\,Jy.
  This large difference can not be explained by epoch or geometry
   differences. However, the dedicated measurement was mis-pointed
  by 24\arcsec\ which caused large uncertainties in the applied
  corrections. The ISOSS flux provides therefore valuable information
  for the colour temperature determination and, through grain size
  models, might give clues whether icy grains were present in the coma
  in December 1997 at almost 4\,AU post-perihelion.
  The last measurement of Method~3a (87380400) was obtained when
  \object{Hale-Bopp} was already 4.9\,AU from the sun. The
  calibrated and reduced 23\arcsec\ flux was 1.97$\pm$0.39\,Jy.
  This is the most distant thermal far-IR observation of \object{Hale-Bopp}
  post-perihelion. A comparison of the flux with a dedicated
  observation (Gr\"un et al. \cite{gruen01}; $F_{\nu}=1.06$\,Jy) at
  a similar distance from the sun, but pre-perihelion, shows that the
  dust emission post-perihelion was higher as the comet receeded from
  the sun. In fact, the higher far-IR fluxes post-perihelion are
  related to contributions from large particles which have been
  accumulated during the passage around the sun and which stay on
  similar orbits as the nucleus.

  Two measurements (32280200, 86880300) show signal peaks a few arcminutes
  away from the nucleus in anti-orbital velocity (trail) direction.
  It seems that the emitting dust particles are not homogeneously
  distributed and are concentrated in a narrow region of the outer
  parts of the dust coma towards the trail direction.
  These features are not seen in slews over other parts of the outer coma.
  Reach et al. \cite{reach00} observed for the first time the dust
  trail formation in comet \object{Encke} in the mid-IR. The ISOSS
  data provide now evidence for this process in the far-IR where
  the emission is strongly connected to the largest particles.
  
\section{Conclusions and Outlook}
\label{sec:con}

  The purpose of the SSO extraction from the ISOSS was manifold:
  Calibration aspects, catalogue cleaning aspects and scientific
  aspects. The main achievements were clearly in the calibration
  section, were serendipitously seen asteroids and planets led
  to an improved flux calibration for ISOSS targets.
  Bright sources of the automatic point-source extraction procedure
  have now a solid calibration basis. It was also possible to establish
  new methods to calibrate source detections under a large variety of
  circumstances, including the important slew end positions and low
  slewing speeds.
  
  The aspect of SSO cleaning from ISOSS catalogue lists
  will avoid wrong identifications and help follow up programmes
  of galactic and extra-galactic sources.
  
  The outcome of the scientific analysis of SSO detections were modest
  due to the limitations of the ISOSS mentioned in Sect.~\ref{sec:scires}.
  Despite all difficulties we could demonstrate that the far-IR fluxes
  of asteroids are important. Diameter and albedo estimates through
  TPM calculations are much more reliable than estimates based on
  visible brightness alone. An accurate H-value of 12.0\,mag would
  allow for diameters ranging from 10.6\,km ($p_V=0.25$) to 23.7\,km
  ($p_V=0.05$), corresponding to a $\pm$40\,\% uncertainty of the
  average. An additional thermal flux with a flux error of $\pm$20\,\%
  allows a 4 times more accurate diameter determination.
  
  The ISOSS results for \object{Hale-Bopp} are more valuable.
  They can now be used for additional comet modeling
  (e.g.\ models by Hanner \cite{hanner83}) for more
  reliable interpretation of grain properties and ice influences at
  different heliocentric distances. Comets are usually
  optically bright due to fresh ice surfaces, but in the
  far-IR the sublimated larger particles
  dominate the thermal emission. After many orbits around the sun these
  large particles form trails which were first measured by IRAS
  (Sykes \cite{sykes86}). For \object{Hale-Bopp} we found significantly more
  thermal emission post-perihelion than for comparable configurations
  pre-perihelion. Additionally we saw asymmetries due to the dust tail
  and an indicative detection of large particles concentrated towards
  the anti-orbital velocity (trail) direction.

  The expectations for future far-IR and submillimetre projects on
  SSO related topics are large: SIRTF, SOFIA, ASTRO-F, HERSCHEL
  and others will have many dedicated programmes on asteroids,
  comets and planets, but will also see by chance interesting targets.
  Especially the ASTRO-F/FIS all sky survey in 4 photometric bands
  in the region 50 to 200\,$\mu$m will serendipitously detect
  many SSOs. Our experience with ISOSS in terms of calibration
  through asteroids and planets, but also in identification of 
  moving targets could then be of great benefit.

\begin{acknowledgements}
  We would like to thank Elwood C. Downey for many helpful discussions
  and for providing {\it xephemdbd} for position calculation of SSOs.
  
  The orbital elements were provided by Gareth Williams, IAU Minor Planet
  Center (http://cfa-www.harvard.edu/iau/mpc.html), for 4 epochs
  during the ISO mission: 1996 Apr.\ 27, 1996 Nov.\ 13, 1997 Jun.\ 1 and
  1997 Dec.\ 18. A modified version of the Uppsala N-body ephemeris 
  programme was used for the final analysis of the pointing accuracy
  between slew and SSO position.
  
  Thanks also to Eberhard Gr\"un, Martha Hanner and Michael M\"uller
  who supported the comet analysis and interpretation and to Dietrich
  Lemke for many valuable comments.
  
  The project was conducted at the ISOPHOT Data Centre, Max-Planck-Institut
  f\"ur Astronomie, Heidelberg, Germany and at the ISO Data Centre,
  Villafranca del Castillo, Spain.
  This project was supported by Deut\-sches Zentrum f\"ur Luft- und
  Raumfahrt e.\,V.~(DLR) with funds of Bundesministerium f\"ur Bildung
  und Forschung, grant no.~50\,QI\,9801\,3.
\end{acknowledgements}

\end{document}